\documentclass{article}

 \usepackage[eandd, final]{neurips_2026}

\usepackage[utf8]{inputenc} 
\usepackage[T1]{fontenc}    
\usepackage{hyperref}       
\usepackage{url}            
\usepackage{booktabs}       
\usepackage{amsfonts}       
\usepackage{amsmath}
\usepackage{nicefrac}       
\usepackage{microtype}      
\usepackage{xcolor}         
\usepackage{enumitem}
\usepackage{multirow}
\usepackage{array}
\usepackage{graphicx}
\usepackage{subcaption}
\usepackage{fancyvrb}
\usepackage{xspace}

\newcommand{\eg}{e.\,g.\xspace}
\newcommand{\name}{KuaiRecV2}

\title{\name: Benchmarking Large-Scale Continual Learning for Diversified and Multi-task Recommendation}

\newcommand{\USTC}{University of Science and Technology of China}
\newcommand{\KS}{Kuaishou Technology, Beijing, China}

\author{
  Chenxu Li\textsuperscript{\ddag}\thanks{Equal Contributions.}~~
  Shuchang Liu\textsuperscript{\S}\textsuperscript{$\ast$}\thanks{Corresponding authors.}~~
  Hantao Shu\textsuperscript{\S}\AND
  Wei Yuan\textsuperscript{\S}~~
  Zhe Xu\textsuperscript{\ddag}
  Guoqing Hu\textsuperscript{\ddag}
  Yan Wang\textsuperscript{\S}~~
  Biao Yang\textsuperscript{\S}~~
  Xiang Li\textsuperscript{\S}~~
  Cheng Ling\textsuperscript{\S}\AND
  Fan Yang\textsuperscript{\S}~~
  Yongqi Liu\textsuperscript{\S}~~
  Lantao Hu\textsuperscript{\S}~~
  Han Li\textsuperscript{\S}~~
  Kaiqiao Zhan\textsuperscript{\S}~~
  Kun Gai\textsuperscript{\S}\AND
  An Zhang\textsuperscript{\ddag}~~
  Xiang Wang\textsuperscript{\ddag}\\[8pt]
  \textsuperscript{\ddag}\USTC \\
  \textsuperscript{\S}\KS \\
  \texttt{\{cxli2024\}@mail.ustc.edu.cn},\texttt{\{liushuchang\}@kuaishou.com} \\
  \texttt{\{shuhantao,yuanwei05,yangbiao,lixiang44,lingcheng,yangfan}\\
  \texttt{liuyongqi,hulantao,lihan08,zhankaiqiao\}@kuaishou.com}\\
  \texttt{\{xz2002,HugoChinn,an\_zhang\}@ustc.edu.cn},\texttt{\{mrwangyan98,xiangwang1223\}@gmail.com} \\
  \texttt{gai.kun@qq.com}
}

\begin{document}

\maketitle

\begin{abstract}
Recommender systems, which retrieve items of interest to users based on their interaction histories, have recently witnessed several key technological advancements, benefiting from the semantic representation space and model-data scaling.
Nevertheless, practitioners and researchers have frequently encountered critical inconsistencies between simple offline evaluations and complex online results, and this gap still exists in the era of large models.
In this work, we address several key factors that contribute to this inconsistency:
1) the \textit{large-scale continual learning challenge} that assumes a practical demand to maintain long-term model effectiveness under dynamic item pools and distributional shifts;
2) the \textit{combinatorial diversity challenge} that requires the generation strategy to solve the accuracy-diversity trade-off in both the finer token level and the holistic item level; and
3) the \textit{multi-task reward balancing challenge}, which is particularly critical for modern generative models that can accurately model the sequence distribution but are less controllable for the multi-task demands.

To address these challenges and further pave the way for the development of more realistic recommendation solutions, we introduce \name, a deliberately constructed benchmark with industrial-level datasets from Kuaishou.
Specifically, we provide a large-scale interaction dataset with million-scale multi-modal item information and long-term interaction records across 30 days, with a dynamic video candidate pool.
Additionally, we provide a rigorous benchmark that evaluates retrieval accuracy, diversity, and multi-task performance in one rubric system, where an effective advancement occurs if and only if the new method improves all metrics.
The code is available at \url{https://github.com/cxli2333/KuaiRecV2}.
\end{abstract}

\section{Introduction}\label{sec: introduction}

Recommender systems~\cite{aggarwal2016recommender,francesco2022recsys} capture user characteristics and preferences from their interaction histories and filter new content of interest that fulfills users' demands for daily information consumption.
Due to the similarity in the sequential modeling formulation, recommendation technologies have witnessed significant development in expressiveness and accuracy, as they benefit from the advancements of large language models and multi-modal models~\cite{zihuai2024llmrssurvey,jinghao2025benefitFromLLM}.
In addition to the LLM-enhanced recommendation paradigm~\cite{yu2025tagcf}, recent discussions on the potential of LLM-style recommendation~\cite{liu2025outshine,yashar2024genrecReview} also reveal that scaling the sequential recommender on generalized semantic representations may achieve state-of-the-art performance, with generative retrievers~\cite{rajput2023tiger,zhai2024hstu,guorui2025onerec,xiaoxi2025genIRsurvey,hou2026bridging} and generative rankers~\cite{shuchang2023gfn4list,huang2025genrank,zhang2026goalrank,jia2026local} as representatives.
The great expectation under these new methodologies is a more guaranteed performance boost with the scaling of data and model size, i.e., achieving a better scaling law behavior compared with conventional recommenders~\cite{zhang2026goalrank}.

\begin{figure}[t]
    \centering
    \includegraphics[width=\linewidth]{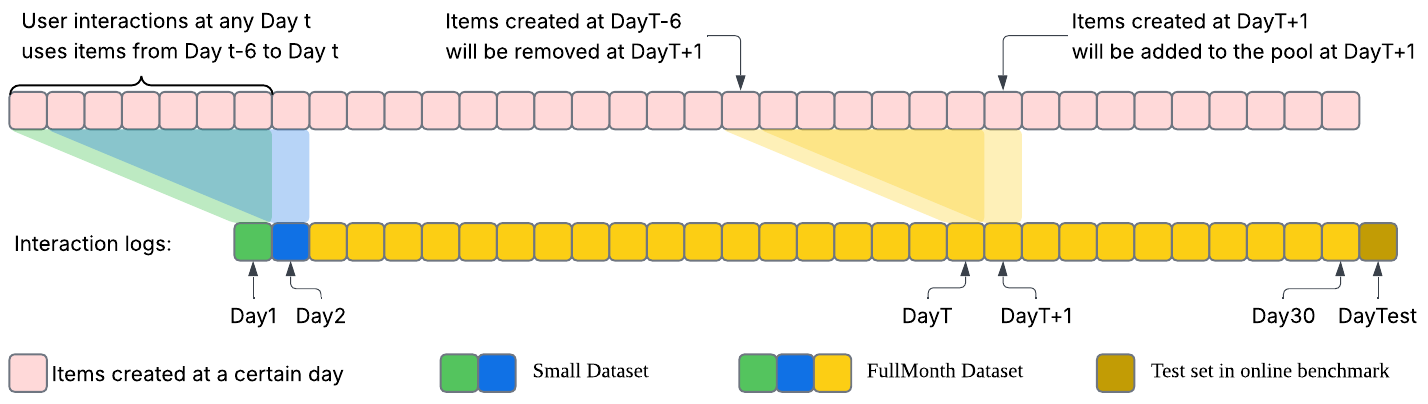}
    \caption{Data Organization of \name.}
    \label{fig: data_organization}
\end{figure}

\subsection{Practical Concerns for LLM-style Recommenders}\label{sec: introduction_concerns}

Technically, these LLM-style generative models typically consist of: a) an \textit{item encoder}, e.g., the item tokenizer, that transforms the item information into a token sequence in a unified representation space; b) a \textit{model} with scalable sequential recommender architecture that captures the sequential patterns in the unified token space; c) a \textit{decoder}, i.e., an item retriever that outputs items based on the generated tokens from the recommender; and d) a \textit{trainer} that either applies generative, discriminative, or hybrid learning strategies.
Though promising, as this field has matured from its early stage,
designers have frequently encountered critical challenges in practice: 
\begin{itemize}[leftmargin=*]
    \item \textbf{Continual Learning under Dynamic Content Pool (RQ1):} In content-intensive scenarios, e.g., video sharing platforms, composers frequently upload new materials and consumers continuously interact with the platform on a daily basis.
    On the one hand, this results in a highly dynamic item pool that forces the recommender system to design robust and long-term consistent solutions.
    Some studies~\cite{geng2022p5,wang2024letter,li2025survey} show that generative recommenders with generalized item representation methods (e.g., semantic representations) are more effective at handling incoming new items, but others notice that item IDs can become more expressive after these items receive sufficient interactions~\cite{zhu2026gatesid,TS-Rec:jiaweifeng,chen2024treatment}.
    On the other hand, samples are chronologically fed into the training process in practice, following a continual learning environment~\cite{bidaki2025continuallearning} that directly avoids the pitfall of ``learning the future, evaluating the past''( which is ubiquitous in conventional benchmarks~\cite{gao2022kuairec,fu2025forge}).
    This means that we need a more rigorous evaluation-while-training protocol for recommender comparisons, but it also brings new problems, including catastrophic forgetting~\cite{aleixo2024catastrophicforgetting}, robustness under distributional shifts~\cite{wan2024onlinedrift}, and restricted computational resources~\cite{aljundi2019gradient}.
    \item \textbf{Controllable Combinatorial Diversity in Token Space (RQ2)}: 
    The item generation involves a token generation phase and a token-to-item retrieval phase, where both phases could be significantly biased even if the tokens are well-separated in the semantic space. 
    In other words, the challenge of the accuracy-diversity trade-off~\cite{jaime1998mmr,ziegler2005intralistdiversity,zhao2025diversitysurvey}, which is widely studied with respect to items, also exists at the token level. 
    In some extreme cases, this problem could be even worse since each item is typically represented by multiple tokens, where token-wise biases are inevitably aggregated.
    Consequently, solving this trade-off is expected to be much more challenging at the token level since the diversity-boosting strategy has to consider the combinatorial effect (e.g., using diversified beam search).
    On the other hand, diversity demands are often induced from explicit business requirements that have predefined categories. 
    This terminology is not necessarily the same as the tokenization method in LLM-style generative recommenders; Thus, more effort is needed to align the diversity control protocols.
    \item \textbf{Multi-task Balancing and Reward Design (RQ3):} Different from discriminative recommenders that can directly formulate the multi-task problem with multi-head/multi-expert predictive models~\cite{ma2018moe,tang2020ple,wang2023mtlrecsurvey,zijian2024m3oe,chen2025dlcrec}, generative sequential recommenders are not directly designed to optimize and control different metrics.
    During training, some studies show that multi-task SFT may help improve the generalizability of the model~\cite{liu2025onerecthink,yu2026sigma,liu2026unity}, but the computational cost becomes a bottleneck due to the duplicated runs for each task during inference.
    Other solutions adopt an additional RL-based preference alignment phase~\cite{guorui2025onerec,yu2026sigma,DBLP:journals/corr/abs-2602-00632} that merges all objectives into a reward model for generator training guidance or inference-time reranking.
    Yet, there is no evidence that the current design of the reward is optimal for multi-task settings, neither is there a conclusion on whether training with multi-task learning or training with a single reward is a better strategy.
\end{itemize}

\subsection{Contributions:}\label{sec: introduction_contribution}

To address these challenges and further investigate the effectiveness of modern recommendation solutions in a more realistic environment, we introduce \name, a benchmark that consists of deliberately constructed industrial-level datasets, rigorously augmented evaluation protocols, and a standardized leaderboard.
The data were collected from the social video-sharing platform Kuaishou~\footnote{www.kuaishou.com}.
Correspondingly, we summarize our contributions as follows:

\begin{itemize}[leftmargin=*]
    \item \textbf{Large-scale industrial datasets:} 
    We first provide the \name-FullMonth dataset that consists of 124k users' interaction records over 30 days, to answer the large-scale continual learning challenge.
    The interactions on the last day are held out for leaderboard evaluation.
    We further select the first two days of the interactions as \name-Small dataset, to recover standard offline evaluation and enable quick model validity verifications.
    For all datasets, we provide multi-modal features on items (including multi-modal embeddings and semantic IDs).
    We summarize the organization of the datasets in Figure \ref{fig: data_organization} and the main statistics in Table \ref{tab: datasets}.
    \item \textbf{The long-term recommendation pipeline:} With the FullMonth dataset, we formulate the \textit{long-term recommendation problem}, where the solution models are required to handle the dynamic first-in, first-out video candidate pool and evolving user interests.
    This environment also addresses the importance of model stability under continual learning strategies, which favor a generalized tokenizer and a balanced recommendation policy between cold-start items and hot items.
    One counterintuitive result we observe is that the RL-based alignment without restriction may potentially downgrade the model performance in the long term.
    \item \textbf{Augmented evaluation:} The evaluation protocol augments the conventional learning-to-rank metrics with multi-level diversity estimation and multi-task evaluation.
    We empirically show that existing generative recommenders may suffer from challenges such as catastrophic forgetting, unstable long-term performance, diversity bottlenecks, and multi-task trade-offs.
\end{itemize}
\section{Datasets and Recommendation Pipeline}\label{sec: data}

\subsection{Data Preparation}\label{sec: data_collection}

\textbf{Data Collection:} Our \name~datasets are collected from Kuaishou, a video-sharing platform serving hundreds of millions of users every day with a billion-scale candidate item set. We uniformly sample users from the full user population and collect their interaction logs over 30 days. We then apply user-side n-core filtering~\cite{he2016vbpr} with n=100, retaining users with at least 100 interactions in the collection window to ensure sufficiently long histories for sequential-model evaluation. For each day, we retain interactions involving items created within the preceding seven days, forming a dynamic candidate item pool.
As summarized in Figure \ref{fig: data_organization}, we provide two benchmark datasets: \name-FullMonth and \name-Small. The FullMonth dataset also includes a hold-out test set for the online leaderboard~\footnote{https://huggingface.co/spaces/cxcx2333/KuaiRecV2Leaderboard}. Table \ref{tab: datasets} summarizes their key statistics, and Appendix \ref{app: data_description} provides additional analysis.

\textbf{Data Structure:} Each user-item interaction record contains labels for multiple tasks, including viewing behaviors (e.g., short view and complete view), positive interactions (e.g., like and comment), and a negative interaction label (hate). For each retained item, we provide a 64-dimensional multimodal embedding using the model in \cite{li-etal-2026-compressing}, an eight-level semantic ID generated by a fixed RQ-VAE-style tokenizer~\cite{rajput2023tiger}, and MLLM-inferred tag IDs~\cite{wu2026guider} for semantic matching analysis. All these models were fixed during the generation of item meta information, and none was trained on the released 30-day interaction logs or held-out interactions, to avoid information leakage. The tokenizer card reports its training corpus, time window, and model version, and separately discloses any item-content overlap with external pretraining data. User IDs, item IDs, and tags are independently reindexed within the released datasets, and user profile features are omitted for privacy protection. Detailed data descriptions and sample examples are provided in Appendix \ref{app: data_description}.

\begin{table}[t]
    \centering
    \begin{tabular}{c|c|c|c|c|c|c}
        \toprule
        subset & \#user & \#item & \#records & sparsity & \#item/user & \#user/item \\
        \midrule
        FullMonth & 124,055 & 24,898,648 & 484,045,145 & 99.9843\% & 3901.85 & 19.44 \\
        Small & 124,054 & 2,452,987 & 38,955,949 & 99.9871\% & 314.02 & 15.88 \\
        \bottomrule
    \end{tabular}
    \caption{Statistics of KuaiRecV2 datasets.}
    \label{tab: datasets}
\end{table}

\begin{figure}[t]
    \centering
    \includegraphics[width=\linewidth]{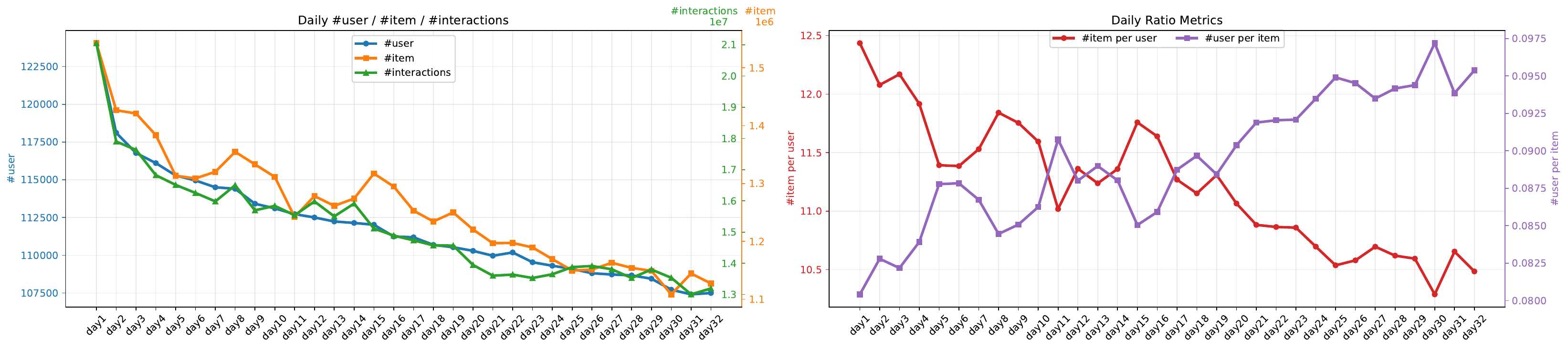}
    \caption{Daily Statstics of \name.}
    \label{fig: daily_stat}
\end{figure}

\subsection{The Long-term Recommendation Pipeline}\label{sec: data_pipeline}

In addition to the \name-FullMonth dataset, we provide a recommendation pipeline designed for a more realistic continual learning environment.
In such an environment, the recommendation models are required to adapt to the time-varying item candidate pool and dynamic interaction sequences.
Specifically, we summarize the following key features:
\begin{itemize}[leftmargin=*]
    \item \emph{Consecutive training on a daily basis:} On the first day, the pipeline trains a randomly initialized generative recommender or a pretrained model using the data from Day 1, i.e., $\mathcal{D}_{1}$. Denote the resulting model as $\theta_1$. Starting from $t=2$, for each subsequent day, the model parameters $\theta_{t-1}$ learned from day $t-1$ are used to initialize the training on the newly arrived data $\mathcal{D}_t$, generating a new model $\theta_t$. In this way, the model is continually updated as the data stream evolves over time.
    \item \emph{Cross-day history context:} To preserve temporal consistency across days, for each user, when constructing the first few interaction instances of a new day, we concatenate the last several interactions from the previous day as historical context. 
    This design avoids breaking user sequences at day boundaries, providing a more accurate representation of real-world streaming recommendation scenarios.
    For better regularization of the history lengths, we provide hyperparameter controls on their minimum $L_\text{min}$ and maximum $L_\text{max}$.
    \item \emph{Dynamic candidate pool: } As illustrated in Figure \ref{fig: data_organization}, we use a sliding-window strategy that only includes items created within the past 7 days for each day's interactions. For any Day $t$, the candidate pool will remove the candidate items created at Day $t-7$ and add items created at Day $t$.
    As we will show in section \ref{sec: benchmark_result_dynamic_item_pool}, one may observe a downgrade of model performance whenever a new day's data comes in with unseen items, even if they are using the shared token space.
    This addresses the need for stable model training and inference under a continual learning environment.
    In practice, the item removals and increments may happen at any time of the day, inducing smoother shifts in the distribution.
    Yet, the challenge of the model stability under candidate pool changes in long-term recommendation scenarios still exists.
\end{itemize}

\section{Benchmarks}\label{sec: benchmark_setup}

Though the pipeline in section \ref{sec: data_pipeline} leaves specific model and training settings to researchers and future studies, we provide implementations of several key baselines and basic training strategies,
in addition to the pipeline and the augmented evaluation protocol.

\subsection{Task Formulation and Augmented Evaluation Metrics}\label{sec: benchmark_augmented_metrics}

\textbf{Evaluation Protocol:} In the long-term recommendation pipeline, we implement a daily leave-one-out evaluation, i.e., for each $\mathcal{D}_t$ and any given user's interaction logs in $\mathcal{D}_t$, all interactions before the last one (of the user) are used for training, while the final one is used for validation. Nevertheless, we encourage readers to customize their evaluation protocols for more advanced settings.
Note that for the FullMonth data, we engage in continual learning through $\{D_1, D_2, \dots, D_{30}\}$, where $D_i$ represents the data of the $i$-th day.
For each day t, we first compute every metric on the $N_t$ validation instances of that day. When a single FullMonth score is required, we use the instance-weighted aggregation $M_{FullMonth} = (\sum_{t}{N_tM_t} )/(\sum{N_t})$, where $M_t$ is the corresponding daily metric. We retain the daily scores as the primary report because the trajectory itself exposes adaptation and forgetting. For backward evaluation, checkpoint $\theta_k$ is evaluated on the Day 1 instances, and the resulting score is reported as the Day-k-to-Day-1 retention performance.

\textbf{Problem Formulation:} 
Given a user request that consists of a user ID and the interaction history $\mathcal{H}$, the recommendation model $\theta$ is responsible for retrieving the item candidates that best align with user preferences.
Formally, in addition to the user history, each training sample includes a target item $i$ and multi-behavior feedback $Y_i = (y_{i,1},\dots,y_{i,k})$ for $k$ labels on this item.
Then the goal of the optimization includes \textbf{three major metric families}:
\begin{itemize}[leftmargin=*]
    \item \emph{Retrieval accuracy metrics:} Following standard evaluation for retrievers, include ranking metrics such as \textbf{NDCG} and \textbf{HitRate (HR)} for the main behavior ``click''. Note that under a continual learning environment, we aim to learn a model that exhibits stability on these metrics across training days or even gradually increases these metrics with the long-term recommendation pipeline.
    \item \emph{Multi-level diversity metrics:} We first provide the conventional item-level diversity metrics such as \textbf{Item Coverage (IC)} and \textbf{GINI} coefficient.
    To better observe the combinatorial effects on the diversity, we include several token-level diversity metrics:
    \textbf{TopK Layer Diversity} reflects diversity in top-K predictions, \textbf{Average Path Diversity} measures structural diversity of recommendation paths, \textbf{Average Layer Unique Tokens} quantifies token variety, and \textbf{Average Layer Entropy} captures prediction uncertainty. Detailed explanations are provided in the appendix \ref{app: detail_diversity}.
    \item \emph{Multi-task metrics:} For the $k$ labels in $Y$ space, we set corresponding rate-prediction tasks as $B=(b_1,\dots,b_k)$ (e.g., like-rate prediction for behavior ``like'') and observe the ranking metrics (i.e., \textbf{NDCG} and \textbf{HR}) for each of the task.
    For behavior-wise evaluation, we further partition the test instances according to the behavior type of the held-out interaction. 
    We additionally allow the training data to be filtered based on the choice of task, so that the model can perform multi-task SFT.
    Nevertheless, we enforce a single-retrieval evaluate-all strategy during evaluation, where only one model can provide recommendations for the estimation of all task metrics.
\end{itemize}

\subsection{Implemented Baselines}\label{sec: benchmark_baselines}

\subsubsection{Models}

For reproduction purposes and to identify the challenges of generative sequential recommenders (mentioned in section \ref{sec: introduction}), we implemented several key baselines under the sequential recommendation formulation: \textbf{SASRec}~\cite{kang2018selfattentivesequentialrecommendation}, \textbf{GRURec}~\cite{hidasi2016sessionbasedrecommendationsrecurrentneural},
\textbf{TIGER}~\cite{rajput2023recommendersystemsgenerativeretrieval},
\textbf{LLaDA-Rec}~\cite{shi2025lladarecdiscretediffusionparallel},
\textbf{MiniOneRec}~\cite{kong2025minionerecopensourceframeworkscaling}.
More detailed information about these models can be seen in Appendix \ref{app:exp_models}.

\subsubsection{Training Strategies}\label{sec: benchmark_training}

\textbf{Task-aware sequence modeling:} 
Given the set of tasks $B$ that correspond to $k$ types of interactions,
while our default setting uses ``click'' sequences for model training and evaluation, we remind readers that constructing user interaction sequences based on different behavioral types in $B$ may better suit the learning of the corresponding task~\cite{guorui2025onerec}. The comparison for this part is in the appendix \ref{app: compare_trainingdata}.

\textbf{SFT vs. RL:} In addition to standard SFT on the next-item prediction dataset, we extend the recommendation model with reinforcement learning (RL) to optimize recommendation lists toward user feedback-oriented objectives. 
Given a user history $\mathcal{H}$, the policy $\theta$ generates a ranked list of items $\mathcal{X}$.
To align with user preferences, we design reward functions, formatted as $R_B = \sum_{b \in B} \alpha_b R_b$, based on behavior types in $B$, and different interaction behaviors are given different levels of significance $\alpha$ in user preference estimation. 
This enables the RL policy to favor recommendations that are more likely to induce desired user responses.
Note that this reward function is implemented merely as training signals but not evaluation metrics, and we evaluate the RL-optimized model under the same protocol as in section \ref{sec: benchmark_augmented_metrics}.

\subsection{Related Benchmarks}\label{sec: related_work_surveys}

Table~\ref{tab: related_benchmarks} compares KuaiRecV2 with related benchmarks. Existing datasets such as FORGE~\cite{fu2025forge} and MUSE~\cite{wu2025musesimpleeffectivemultimodal} are mainly designed for SID quality assessment and user interest modeling, respectively, while KuaiRec~\cite{gao2022kuairec} is an earlier version of the benchmark that focuses on fully-observed dataset evaluation in a much smaller-scale setting. 
In contrast, \name~is built for continual learning, combinatorial diversity, and multi-task retrieval.
The substantially large-scale dataset enables a better scaling law observation for existing methods. It also adopts extracted SID and Multi-Media Understanding (MMU) embeddings to support diversified generative recommendation modeling and evaluation.
Full related work is provided in Appendix \ref{sec: related_work}.

\begin{table*}[t]
    \centering
    \begin{tabular}{c|c|c|c}
        \toprule
         & Representation & Scale & Tasks \\
        \midrule
        \parbox{0.15\textwidth}{FORGE~\cite{fu2025forge} \\(2025.09)} &\parbox{0.2\textwidth}{Extracted SID \& MMU embeddings} & \parbox{0.2\textwidth}{14B interactions, 100K users, 250M items} & SID quality assessment \\
        \midrule
       \parbox{0.15\textwidth}{MUSE~\cite{wu2025musesimpleeffectivemultimodal}\\(2025.12)} &\parbox{0.2\textwidth}{User/Item feature \& MMU embeddings} & \parbox{0.2\textwidth}{99M interactions, 8M users, 35M items} &User Interest Modeling \\
        \midrule
        \parbox{0.15\textwidth}{KuaiRec~\cite{gao2022kuairec}\\ (2022.02) }&\parbox{0.2\textwidth}{User/Item feature} &\parbox{0.2\textwidth}{4M interactions, 1,411 users, 3,327 items} & \parbox{0.2\textwidth}{Fully-observed Dataset Evaluation} \\
        \midrule
        \parbox{0.15\textwidth}{\name} & \parbox{0.2\textwidth}{Extracted SID \& MMU embeddings} & \parbox{0.2\textwidth}{484M interactions, 124K users, 24M items} & \parbox{0.2\textwidth}{Long-term Recommendation\& Diversity\&
        Multi-task Retrieval} \\
        \bottomrule
    \end{tabular}
    \caption{Comparisons between related benchmarks.}
    \label{tab: related_benchmarks}
\end{table*}

\section{Benchmark Results, Key Observations, and Open Questions}\label{sec: benchmark_result}

\subsection{Evaluating Continual Adaptation to Dynamic Item Poolsol (RQ1) }\label{sec: benchmark_result_dynamic_item_pool}

\begin{figure}[t]
    \centering

    \begin{minipage}[t]{0.57\textwidth}
        \centering
        \includegraphics[width=\linewidth]{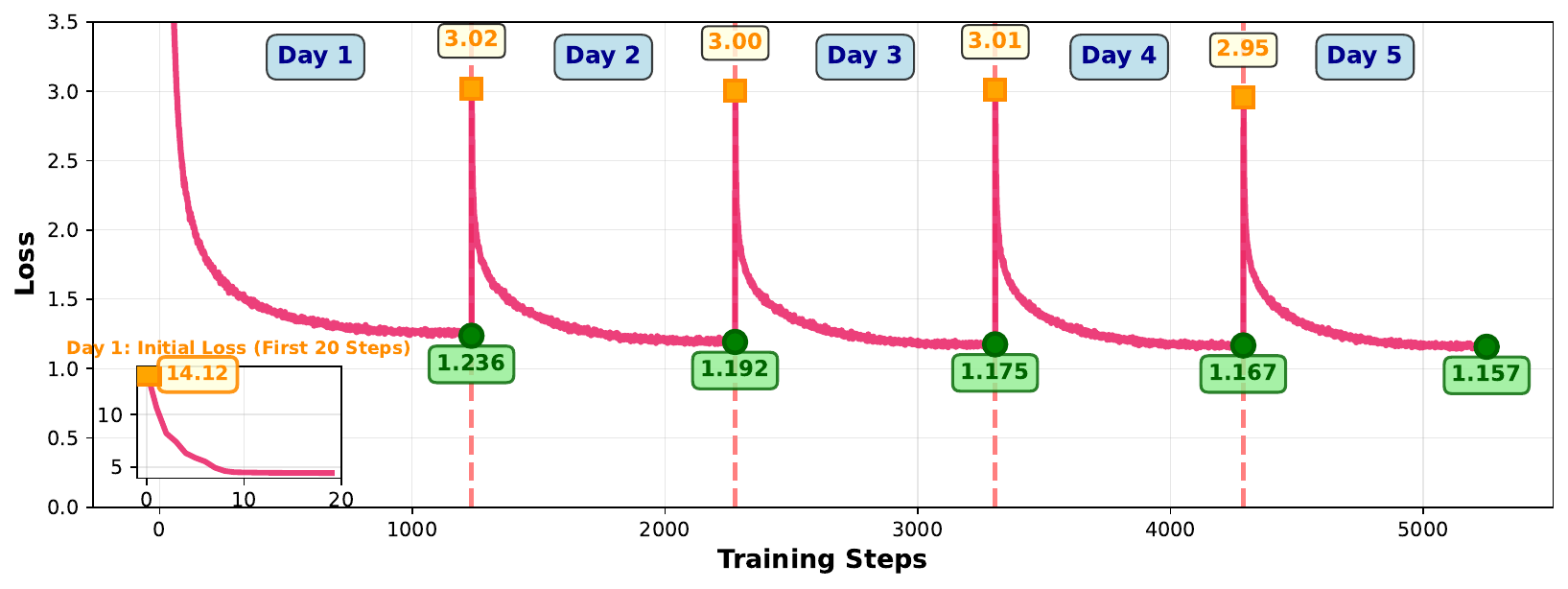}
        \captionof{figure}{Training loss during 5-day continual learning. Orange squares indicate the initial loss at each day boundary (when item pools change), and green dots show the final loss after training. The inset shows the rapid initial convergence on Day 1.}
        \label{fig:loss}
    \end{minipage}
    \hfill
    \begin{minipage}[t]{0.39\textwidth}
        \centering
        \includegraphics[width=\linewidth]{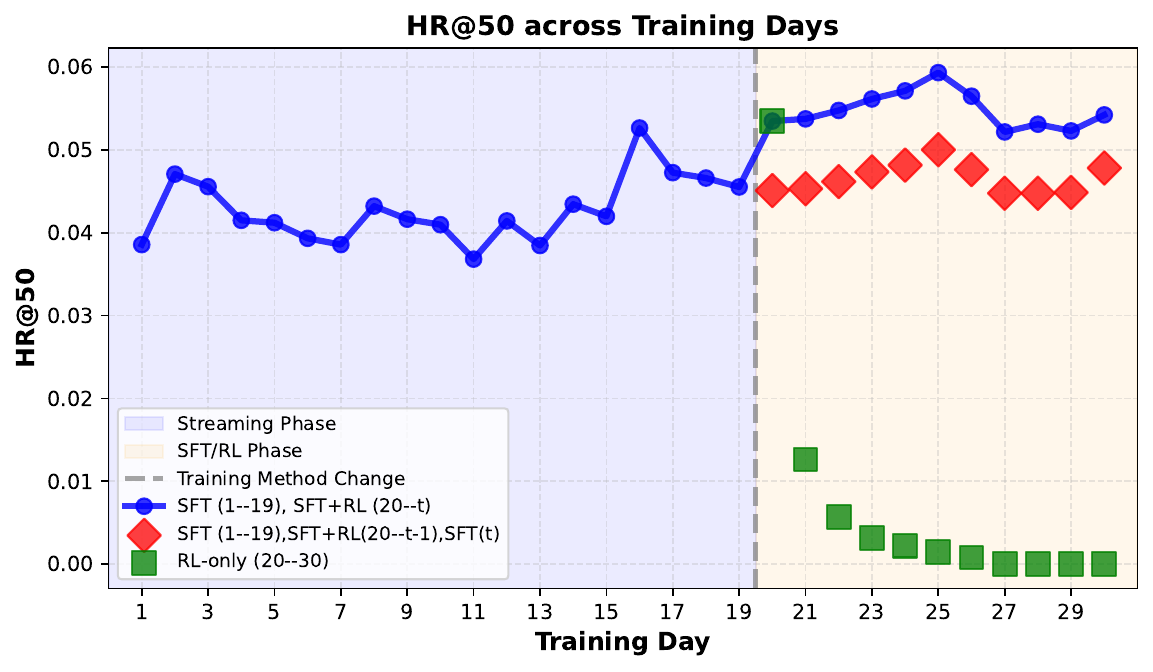}
        \caption{HR@50 across training days. We compare the results of alternating SFT and RL with those of RL-only.}
        \label{fig:hr50}
    \end{minipage}

\end{figure}

We evaluate generative models on \name-FullMonth with evolving item pools using the long-term recommendation pipeline (section \ref{sec: data_pipeline}). 
Figure \ref{fig:loss} presents the loss curve of MiniOneRec~\cite{kong2025minionerecopensourceframeworkscaling} during training over the first five days on the FullMonth dataset. 
It indicates that item replacement under a dynamic item pool requires the model to relearn and adapt continuously. More detailed results on the full 30-day continual learning are provided in Appendix \ref{app: detail_results}.

\textbf{The existence of distributional shift and catastrophic forgetting in long-term recommendation.} 

We evaluate backward transfer by testing each checkpoint on the Day 1 validation instances after training through Day k. We compare MiniOneRec with two continual-learning baselines: EWC LoRA, which applies elastic weight consolidation to LoRA parameters during daily updates, and Replay, which retains 5\% of previously observed training instances in a replay buffer. Table \ref{tab:back to day1} reports Day-1 HR@50. EWC LoRA improves retention through Day 5, while Replay performs best from Day 2 onward, reaching 3.20 after Day 7. Despite these gains, all methods show substantial forgetting, highlighting the need to evaluate both adaptation to the current item pool and retention of earlier preferences.

\begin{table*}[t]
\centering
\small
\caption{Backward transfer (HR@50 on Day 1), checkpoints trained Days 1--7.}
\label{tab:back to day1}
\begin{tabular}{lrrrrrrr}
\toprule
Model & Day 1 & Day 2 & Day 3 & Day 4 & Day 5 & Day 6 & Day 7 \\
\midrule
MiniOneRec & 39.09 & 17.89 & 6.625 & 1.03 & 0.41 & 0.54 & 0.19 \\
MiniOneRec + EWC LoRA & 38.67 & 22.93 & 7.18 & 1.13 & 0.60 & 0.53 & 0.20 \\
MiniOneRec + Replay (5\%) & 39.09 & 25.24 & 10.26 & 5.59 & 5.45 & 3.97 & 3.20 \\
\bottomrule
\end{tabular}
\end{table*}

\textbf{Item cold-start and temporal dynamics of user interests}: New items enter the catalog daily, but they lack historical interactions, forcing models to balance exploring new items with exploiting known user preferences---a trade-off that is particularly acute in continual learning.
While using semantic representations like SID and MMU embeddings may mitigate the issue, we still observe daily loss spikes in Figure \ref{fig:loss}, indicating insufficient adaptation.
Systematically, the dynamic item pool also contributes to user preference shifts, inducing different patterns at different timescales---short-term interests may shift daily, while long-term preferences remain stable. Current approaches treat all temporal dynamics uniformly, failing to capture the multi-scale nature of user behavior evolution.
Intuitively, we expect a recommender system that can reduce catastrophic forgetting of long-term preferences rather than short-term preferences.

These interconnected challenges highlight why continual recommendation is fundamentally more difficult than static settings and underscore the need for algorithms that can efficiently handle distribution shifts, item cold-start scenarios, and multi-scale temporal dynamics. 

\subsection{Token-level and Item-level Diversity Metrics (RQ2) }



We evaluate all implemented models on \name-Small.
We summarize their ranking performance in Table \ref{tab:ndcg_hr} and their item-level diversity metrics in Table \ref{tab:models_diversity_results}.
Note that due to the excessive number of items, we only calculate the Gini coefficient for items that appear once or more.
More detailed results are provided in the appendix \ref{app: diversity_results}. 
It can be seen from these results that the diversity performance of model recommendations is poor, with a high Gini coefficient value and low item coverage. 
On average, there are only 12 categories at the first level across fifty groups of SID, while users' interests in real-world scenarios are far more diverse than this number. 

\textbf{Combinatorial diversity in the token level:} 
Table \ref {tab: token_diversity_metrics} summarizes the token-level diversity metrics obtained from the MiniOneRec model on training day 21.
We can see that most of the diversity is achieved by the first few tokens of an item, indicating that the latter tokens are extremely biased, with little room for diversity control.
With both the item-level and token-level diversity metrics available, recommender systems are now entering a more sophisticated but flexible state on diversity control.
For example, one can choose whether the given user demands a more accurate recommendation with items in the same first-level token but differing in latter positions, or a more diverse recommendation with items from different first-tokens.

\textbf{Alignment with existing terminologies:}
As we have explained in section \ref{sec: introduction}, diversity demands are often induced from explicit business requirements that have predefined categories.
This terminology is not necessarily the same as the tokenization method in LLM-style generative recommenders; Thus, more efforts are needed to align the diversity control protocols.
Fortunately, although SID is not a manually defined classification and cannot directly match actual business terms, the SID token-level diversity metric can still reflect the variety of item semantics in the language space, partially optimizing business-defined diversities with implicit control.


\begin{table}[htbp]
\centering
\caption{Performance comparison in terms of NDCG and HR at different cutoffs. Values are means over three random seeds (42, 2026, and 999), with standard deviations in parentheses. All values are reported in $10^{-3}$ for readability.}
\label{tab:ndcg_hr}
\begin{tabular}{lccccc}
\toprule
Model & NDCG@1 & NDCG@5 & NDCG@10 & NDCG@20 & NDCG@50 \\
\midrule
SASRec      & 0.28 (0.03) & 0.68 (0.08) & 0.97 (0.11) & 1.31 (0.09) & 1.92 (0.11) \\
GRURec      & 0.31 (0.04) & 0.80 (0.05) & 1.20 (0.08) & 1.71 (0.10) & 2.82 (0.16) \\
LLaDA-Rec   & 0.78 (0.08) & 2.25 (0.10) & 3.10 (0.11) & 4.21 (0.08) & 5.60 (0.11) \\
TIGER-Tiny  & 1.26 (0.09) & 2.96 (0.12) & 4.01 (0.17) & 5.26 (0.31) & 6.98 (0.55) \\
TIGER-Small & 2.12 (0.07) & 5.34 (0.09) & 7.37 (0.17) & 9.81 (0.17) & 12.90 (0.26) \\
MiniOneRec  & 3.53 (0.04) & 8.72 (0.07) & 12.01 (0.15) & 15.33 (0.18) & 19.22 (0.82) \\
\bottomrule
\end{tabular}

\vspace{0.8em}
\begin{tabular}{lccccc}
\toprule
Model & HR@1 & HR@5 & HR@10 & HR@20 & HR@50 \\
\midrule
SASRec      & 0.28 (0.03) & 1.10 (0.08) & 1.96 (0.11) & 3.31 (0.08) & 6.61 (0.10) \\
GRURec      & 0.31 (0.04) & 1.32 (0.06) & 2.71 (0.12) & 5.05 (0.48) & 10.99 (0.62) \\
LLaDA-Rec   & 0.78 (0.08) & 3.71 (0.17) & 6.47 (0.20) & 11.02 (0.29) & 17.31 (0.42) \\
TIGER-Tiny  & 1.26 (0.09) & 4.72 (0.16) & 8.01 (0.70) & 13.17 (1.29) & 20.52 (3.27) \\
TIGER-Small & 2.12 (0.07) & 8.71 (0.12) & 15.58 (0.61) & 25.07 (0.72) & 41.07 (0.81) \\
MiniOneRec  & 3.53 (0.04) & 14.05 (0.12) & 23.59 (0.33) & 38.45 (0.41) & 62.63 (1.10) \\
\bottomrule
\end{tabular}
\end{table}

\begin{table}[htbp]
  \centering
  \caption{Diversity metrics of different models at various cutoffs. All models exhibit limited diversity, as reflected by high Gini values and low coverage scores across cutoffs.}
  \label{tab:models_diversity_results}
  \begin{tabular}{lcccccccccc}
    \toprule
    \multirow{2}{*}{Models} & \multicolumn{5}{c}{Gini ($\times 10^{-2}$)} &\multicolumn{5}{c}{Coverage ($\times 10^{-4}$)}\\
    \cmidrule(lr){2-6} \cmidrule(lr){7-11}
    & @1 & @5 & @10 & @20 & @50 & @1 & @5 & @10 & @20 & @50\\
    \midrule
    LLaDA-Rec & 91.43 & 92.33 & 93.10 & 93.60 & 92.41 & 1.26 & 4.45 & 9.56 & 23.53 & 98.86 \\
    TIGER-Tiny & 93.69 & 96.07 & 96.42 & 96.47 & 95.87 & 0.83 & 4.23 & 8.69 & 18.42 & 48.73 \\
    TIGER-Small & 86.69 & 90.10 & 91.50 & 92.51 & 93.10 & 1.83 & 5.62 & 10.20 & 20.05 & 54.38 \\
    MiniOneRec & 88.10 & 89.50 & 89.60 & 89.51 & 90.69 & 1.28 & 3.25 & 5.03 & 8.13 & 18.80 \\
    \bottomrule
  \end{tabular}
\end{table}

\begin{table}[htbp]
  \centering
  \caption{Diversity metrics across layers for streaming training day 21 (click behavior).}
  \label{tab: token_diversity_metrics}
  \begin{tabular}{llcccccccc}
    \toprule
    \multicolumn{2}{c}{Metric} & \multicolumn{8}{c}{Layer}\\
    \cmidrule(lr){3-10}
    & & 1 & 2 & 3 & 4 & 5 & 6 & 7 & 8\\
    \midrule
    \multicolumn{2}{l}{Avg Path Diversity} & 12.47 & 31.34 & 42.84 & 45.93 & 46.61 & 46.83 & 47.50 & 48.69 \\
    \multicolumn{2}{l}{Avg Layer Unique Tokens} & 12.47 & 25.61 & 38.64 & 42.81 & 43.53 & 43.84 & 44.86 & 46.04 \\
    \multicolumn{2}{l}{Avg Layer Entropy} & 2.74 & 4.11 & 5.09 & 5.32 & 5.35 & 5.37 & 5.42 & 5.47 \\
    \midrule
    \multirow{5}{*}{\begin{tabular}[c]{@{}l@{}}TopK Layer\\Diversity\end{tabular}} & (Top1) & 1.00 & 1.00 & 1.00 & 1.00 & 1.00 & 1.00 & 1.00 & 1.00 \\
     & (Top5) & 3.66 & 4.55 & 4.74 & 4.77 & 4.77 & 4.79 & 4.78 & 4.80 \\
     & (Top10) & 5.99 & 8.79 & 9.45 & 9.59 & 9.60 & 9.65 & 9.65 & 9.67 \\
     & (Top20) & 8.93 & 15.79 & 18.53 & 19.06 & 19.13 & 19.21 & 19.27 & 19.33 \\
     & (Top50) & 12.47 & 25.61 & 38.64 & 42.81 & 43.53 & 43.84 & 44.86 & 46.04 \\
    \bottomrule
  \end{tabular}
\end{table}

\subsection{Estimating Multi-task Performance (RQ3)}\label{sec: multi-behavior_RL}

We extend the streaming training framework with reinforcement learning to leverage multiple behavior signals. While the SFT stage uses only click sequences, we design RL reward functions based on diverse behaviors (Appendix \ref{app:exp_rl_reward}).
We evaluate the effectiveness of multi-behavior RL by comparing three training strategies starting from Day 20: (1) \textbf{RL-only}, where we apply only RL without SFT; (2) \textbf{SFT-only}, where the model continues to train with SFT and no RL is applied; and (3) \textbf{SFT and RL alternating}, where SFT and RL are performed alternatively in each daily training step.
Figure \ref{fig:hr50} shows the performance comparison across these strategies. Days 1–19 are trained with SFT only. Days 20–30 are trained with alternating SFT and RL. Blue circles denote post-RL results, red diamonds denote pre-RL results, and green markers denote RL-only results.

\textbf{Multi-behavior RL improves performance:} 
On Day 20, the alternating SFT and RL strategy achieves noticeable improvements over the SFT-only baseline, indicating that multi-behavior rewards successfully guide the model toward better performance and user preference alignment.
By first adapting to new data through SFT and then optimizing the objective via RL, this hybrid training scheme better handles the non-stationary nature of streaming recommendation.


\textbf{Data shifting challenges in multi-task recommendation.} 
Compared to SFT+RL and SFT-only methods, the RL-only strategy reveals a critical limitation: although RL improves performance on Day 20, it fails to maintain performance when the item pool changes on Day 21. This sharp degradation suggests that RL alone is insufficient to adapt to new data distributions or large-scale continual learning.
Following these findings, we show evidence that RL-based solutions for multi-task recommendation might be fragile to distribution shifts. 
In our empirical study, multi-tasking RL is effective only when combined with SFT.

\textbf{Task relations and effective advancement.} We further provide Table \ref{tab:behavior_metrics} that presents the performance metrics of MiniOneRec (SFT-only) across different behaviors, evaluated at the 21st training day. 
Semantically related behaviors tend to exhibit similar recommendation patterns. For example, \textit{long view}, \textit{like}, and \textit{profile enter} achieve comparable NDCG and HR values across cutoffs, suggesting that the model captures shared preference structures among positive feedback signals. In contrast, behavior-specific signals can differ substantially. For instance, \textit{hate} remains near zero at all cutoffs, indicating a sharply different retrieval pattern (e.g., anomaly detection) from other behaviors. 
More detailed results are provided in Appendix \ref{app: behavior_results}.
In general, task metrics may also exhibit trade-off performances in practice (e.g., when the model is optimized without careful designs~\cite{zijian2024m3oe}). 
In such cases, we note them as \emph{defective advancement}, and only the new method improves all metrics can be regarded as \emph{effective advancement}.

\begin{table}[htbp]
  \centering
  \caption{Behavior metrics at different cutoffs. Related behaviors show similar ranking performance, while unrelated behaviors, such as \textit{hate}, differ sharply.}
  \label{tab:behavior_metrics}
  \begin{tabular}{lcccccccccc}
    \toprule
    \multirow{2}{*}{Behavior} & \multicolumn{5}{c}{NDCG ($\times 10^{-3}$)} &\multicolumn{5}{c}{HR ($\times 10^{-3}$)}\\
    \cmidrule(lr){2-6} \cmidrule(lr){7-11}
    & @1 & @5 & @10 & @20 & @50 & @1 & @5 & @10 & @20 & @50\\
    \midrule
    click & 2.87 & 7.30 & 9.85 & 12.78 & 16.03& 2.87 & 11.77 & 19.70 & 31.39 & 47.60 \\
    long view & 3.18 & 8.02 & 10.78 & 13.97 & 17.43& 3.18 & 12.89 & 21.49 & 34.20 & 51.50 \\
    like & 2.48 & 7.05 & 9.48 & 12.47 & 16.45 & 2.48 & 12.03 & 19.47 & 31.51 & 50.99 \\
    follow & 0.00 & 7.30 & 9.66 & 10.53 & 10.53& 0.00 & 15.32 & 22.98 & 26.81 & 26.81 \\
    forward & 2.64 & 5.44 & 6.32 & 8.30 & 9.82& 2.64 & 7.91 & 10.55 & 18.46 & 26.38 \\
    comment & 3.74 & 5.20 & 7.29 & 9.38 & 12.72 & 3.74 & 6.54 & 13.08 & 21.49 & 38.31 \\
    search & 0.00 & 3.26 & 10.71 & 14.63 & 16.18 & 0.00 & 7.96 & 31.87 & 47.80 & 55.77 \\
    profile enter & 2.71 & 5.20 & 7.70 & 10.24 & 13.09  & 2.71 & 8.12 & 15.91 & 26.07 & 40.29\\
    hate & 0.00 & 0.00 & 0.00 & 0.00 & 0.00 & 0.00 & 0.00 & 0.00 & 0.00 & 0.00 \\
    \bottomrule
  \end{tabular}
\end{table}

\subsection{Scalability Comparison on Large-Scale Item Pools}\label{sec: benchmark_result_retrieval}
We first evaluate different recommendation architectures on \name-Small. 
This setting allows us to assess how different architectures scale to large item pools before introducing temporal dynamics.
Table \ref{tab:ndcg_hr} shows the detailed results.

\textbf{Generative sequential models substantially outperform traditional sequential models.}
All three generative models (MiniOneRec, LLaDA-Rec, TIGER) achieve higher NDCG compared to traditional models (SASRec, GRU4Rec). 
This gap reflects the scalability limitations of traditional ranking-based models on large item pools. Generative models, designed for large output vocabularies through token-based generation, scale more efficiently.

\textbf{Model capacity matters for generative models.} MiniOneRec (0.5B) and TIGER-Small (30M) substantially outperform smaller models (TIGER-Tiny: 4.9M, LLaDA-Rec: 5.3M), indicating that larger capacity enables better performance on large-scale item pools. Notably, even the smallest generative model substantially outperforms the best traditional model, indicating that architectural choice is more critical than model size alone.

Through our benchmark datasets, we believe that it becomes easier to validate the scaling behavior of existing and future generative recommender solutions, in both static offline and continual learning environments.
\section{Conclusion and Discussion}\label{sec: conclusion}


This paper introduces \textbf{\name}, a benchmark for large-scale diversified and multi-task recommendation under a continual learning environment.
The benchmark covers 124 thousand users, 24 million items, as well as 0.48B interactions across a long-term span of 30 days, providing ultra-long interaction sequences. 
We present the continual learning design choices within our long-term recommendation pipeline, which implements a rigorous evaluation system and an online leaderboard. 
On this basis, \name~narrows the gap between offline experiments and online experiments by a comprehensive evaluation that consists of long-term recommendation accuracy, multi-level diversity, and multi-task balancing.
We present limitations and future work in Appendix~\ref{app:limitations_and_future}.

\section{Acknowledgments}\label{sec: acknowledegments}
This research is supported by the National Natural Science Foundation of China (No.U25A20445 and No.62625608).

\bibliographystyle{unsrt}
\bibliography{main}


\clearpage
\section*{NeurIPS Paper Checklist}


\begin{enumerate}

\item {\bf Claims}
    \item[] Question: Do the main claims made in the abstract and introduction accurately reflect the paper's contributions and scope?
    \item[] Answer: \answerYes{} 
    \item[] Justification: The main claims in the abstract and introduction are consistent with the paper’s theoretical contributions, experimental results, and scope.

\item {\bf Limitations}
    \item[] Question: Does the paper discuss the limitations of the work performed by the authors?
    \item[] Answer: \answerYes{} 
    \item[] Justification: The limitations of this work are explicitly discussed in Appendix \ref{app:limitations_and_future} of the main paper.

\item {\bf Theory assumptions and proofs}
    \item[] Question: For each theoretical result, does the paper provide the full set of assumptions and a complete (and correct) proof?
    \item[] Answer: \answerNA{} 
    \item[] Justification: the paper does not include theoretical results.

    \item {\bf Experimental result reproducibility}
    \item[] Question: Does the paper fully disclose all the information needed to reproduce the main experimental results of the paper to the extent that it affects the main claims and/or conclusions of the paper (regardless of whether the code and data are provided or not)?
    \item[] Answer: \answerYes{} 
    \item[] Justification: We fully disclose all necessary details for reproducing main results, and provide code and data to support faithful reproduction, as detailed in the supplementary material.

\item {\bf Open access to data and code}
    \item[] Question: Does the paper provide open access to the data and code, with sufficient instructions to faithfully reproduce the main experimental results, as described in supplemental material?
    \item[] Answer: \answerYes{} 
    \item[] Justification: We provided code and data.

\item {\bf Experimental setting/details}
    \item[] Question: Does the paper specify all the training and test details (e.g., data splits, hyperparameters, how they were chosen, type of optimizer) necessary to understand the results?
    \item[] Answer: \answerYes{} 
    \item[] Justification: We discuss experimental details in Appendix.

\item {\bf Experiment statistical significance}
    \item[] Question: Does the paper report error bars suitably and correctly defined or other appropriate information about the statistical significance of the experiments?
    \item[] Answer: \answerNo{} 
    \item[] Justification: Statistical significance analysis is not provided in this work. Experimental results are consistent across repeated runs to ensure reliability.

\item {\bf Experiments compute resources}
    \item[] Question: For each experiment, does the paper provide sufficient information on the computer resources (type of compute workers, memory, time of execution) needed to reproduce the experiments?
    \item[] Answer: \answerYes{} 
    \item[] Justification: We have specified the computational resource requirements in the appendix.
    
\item {\bf Code of ethics}
    \item[] Question: Does the research conducted in the paper conform, in every respect, with the NeurIPS Code of Ethics \url{https://neurips.cc/public/EthicsGuidelines}?
    \item[] Answer: \answerYes{} 
    \item[] Justification: We strive to minimize the exposure of any personally identifiable information.

\item {\bf Broader impacts}
    \item[] Question: Does the paper discuss both potential positive societal impacts and negative societal impacts of the work performed?
    \item[] Answer: \answerYes{} 
    \item[] Justification: We present a corresponding discussion in Appendix \ref{app: borader_impacts}.
    
\item {\bf Safeguards}
    \item[] Question: Does the paper describe safeguards that have been put in place for responsible release of data or models that have a high risk for misuse (e.g., pre-trained language models, image generators, or scraped datasets)?
    \item[] Answer: \answerNA{} 
    \item[] Justification: The paper poses no such risks.

\item {\bf Licenses for existing assets}
    \item[] Question: Are the creators or original owners of assets (e.g., code, data, models), used in the paper, properly credited and are the license and terms of use explicitly mentioned and properly respected?
    \item[] Answer: \answerYes{} 
    \item[] Justification: All existing assets (datasets, code, models) are properly credited, and their licenses and terms of use are explicitly stated and respected in the paper.

\item {\bf New assets}
    \item[] Question: Are new assets introduced in the paper well documented and is the documentation provided alongside the assets?
    \item[] Answer: \answerYes{} 
    \item[] Justification: All new assets are well-documented, and the detailed information is provided in both the main paper and the appendix.

\item {\bf Crowdsourcing and research with human subjects}
    \item[] Question: For crowdsourcing experiments and research with human subjects, does the paper include the full text of instructions given to participants and screenshots, if applicable, as well as details about compensation (if any)? 
    \item[] Answer: \answerNA{} 
    \item[] Justification: The paper does not involve crowdsourcing nor research with human subjects.

\item {\bf Institutional review board (IRB) approvals or equivalent for research with human subjects}
    \item[] Question: Does the paper describe potential risks incurred by study participants, whether such risks were disclosed to the subjects, and whether Institutional Review Board (IRB) approvals (or an equivalent approval/review based on the requirements of your country or institution) were obtained?
    \item[] Answer: \answerNA{} 
    \item[] Justification: The paper does not involve crowdsourcing nor research with human subjects.

\item {\bf Declaration of LLM usage}
    \item[] Question: Does the paper describe the usage of LLMs if it is an important, original, or non-standard component of the core methods in this research? Note that if the LLM is used only for writing, editing, or formatting purposes and does \emph{not} impact the core methodology, scientific rigor, or originality of the research, declaration is not required.
    \item[] Answer: \answerYes{} 
    \item[] Justification: We clearly describe the usage of LLMs as an important component in the core methodology of this work, as stated in the main paper.

\end{enumerate}


\clearpage
\appendix
\section{Related Work}\label{sec: related_work}

\subsection{Recommendation Diversity}\label{sec: related_work_diversity}

Besides ranking accuracy metrics, diversity metrics are equally critical in practical recommendation tasks. 
The mainstream applications of diversity indicators are dominated by list-level metrics, including Maximum Marginal Relevance\cite{mmr}, Intra-list Diversity\cite{intra-listdiversity}, Normalized Discounted Cumulative Gain\cite{ndcg}, and Determinantal Point Process Score\cite{DPPS1,DPPS2}. Other works also reveal that system-level diversity metrics such as item coverage\cite{coverage}, entropy\cite{entropy}, GINI-index\cite{gini}, and list variance\cite{listvariance} are also important for maintaining a content-productive environment. In this work, we extend the aforementioned system-level diversity indicators to the token level. For generative recommendation models, token-level diversity metrics can reflect the diversity of generated results in a more granular manner.

\subsection{Generative Recommendation}\label{sec: related_work_genrec}

The pioneering attempts that studied the generative recommendation formulation date back to a decade ago~\cite{wang2017irgan,liang2018multvae,bello2018seq2slate} when deep generative models~\cite{kingma2014vae,goodfellow2014gan,vasvani2017transformer} were proposed in machine learning and later integrated in recommender systems.
The main difference between generative and discriminative methodologies is whether the recommender is modeling the item generation probability (i.e., $P(i|u)$) or the expected feedback from the user (i.e., $P(y|i,u)$).
While the generative modeling of $P(i|u)$ is widely adopted in retrieval tasks~\cite{rajput2023tiger,zhai2024hstu,guorui2025onerec,xiaoxi2025genIRsurvey,hou2026bridging}, the discriminative modeling of $P(y|i,u)$ is more favorable in ranking tasks~\cite{shuchang2023gfn4list,huang2025genrank,jia2026local}.
Despite this divergent viewpoint, both methodologies are functionally feasible for both retrieval and ranking tasks, and some recent works~\cite{shuchang2023gfn4list} also notice that recommenders are fundamentally energy-based models that are simultaneously generative and discriminative (i.e., the generation probability should be proportional or monotonic to the user's positive feedback).

Technically, the generative modeling strategies in recommendation generally consist of two focuses:
1) The first focus is autoregressive generative solutions that aim to model user interaction sequences as next item/list prediction tasks.
Following the insights from the development of large language models(LLMs), new architectures~\cite{rajput2023tiger,guorui2025onerec,guo2025twinflow} are proposed for both retrieval and ranking tasks to reformulate the user interaction sequences with different tokenization techniques.
For list-wise generation, additional designs are required to capture the candidate context information and item correlations in the output list~\cite{mge, gao2025llm4rerankllmbasedautorerankingframework, zhang2026goalrank}.
2) The second focus is non-autoregressive generative solutions~\cite{liu2025preferencediffusionrecommendation,hu2024generateinstantiateprefertextguided,DBLP:journals/corr/abs-2608-24091} that aim to model the distribution of user preferences across all items.

Except for these LLM-style generative sequential recommenders, there are other directions that incorporate generative thinking.
VAE-based models~\cite{kingma2014vae} are capable of generating the entire item space~\cite{liang2018multvae} and list space~\cite{liu2021pivotcvae}, recent efforts have found that diffusion-based models~\cite{ho2020ddpm} and generative flow networks may have higher potential in generating item embeddings~\cite{li2024diffurec}, item scores~\cite{mao2026denoising}, discrete item selection space~\cite{shi2025lladarecdiscretediffusionparallel,hu2026fading}, and list permutation space~\cite{shuchang2023gfn4list,lin2024discrete,chen2026position}.

In this work, our main focus is to study the behavior of modern generative recommendation models on a large scale dataset, especially when the item space is too large to apply standard ID-based retrieval. The proposed \name~benchmark is intentionally made agnostic to the recommendation models.

\subsection{Continual Learning in Recommender Systems}\label{sec: related_work_video_recommendation}

For online systems that receive persistent user engagements and data updates, the notion of continual learning~\cite{wang2024continual,bidaki2025continuallearning} exists as a machine learning problem that addresses the challenges related to system dynamics, such as catastrophic forgetting~\cite{aleixo2024catastrophicforgetting,huang2026generativerecommendersforgetlong}, robustness under distributional shifts~\cite{wan2024onlinedrift}, and restricted computational resources~\cite{aljundi2019gradient}.
Existing continual recommendation methods can be broadly grouped into replay-based\cite{pmlr-v267-wan25d}, regularization-based\cite{DBLP:journals/corr/abs-2506-07466}, and architecture-based approaches\cite{DBLP:journals/corr/abs-2510-25093}. Our work provides the data and fundamental framework for continual recommendation.

\section{Limitations and Future Directions} \label{app:limitations_and_future}
\subsection{Limitations}
While~\name~moves the field towards a more online-alike evaluation of recommender systems, the data used in industrial recommendation systems can still be far more complex than ours (e.g., excluding user features for privacy concerns), making it impossible for us to fully simulate real-world online recommendation systems.
Generally, there is still a major gap between offline and online evaluation of recommender systems.
Besides, we do not consider cross-domain scenarios that allow users to interaction with different recommendation domains that provide a separate set of content, since \name~does not address the generalizability or transferability of recommendation models.

\subsection{Future Directions}

\textbf{Scaling law under continual learning.}
Unlike static benchmarks, streaming recommendation benchmarks feature dynamic iteration of item sets, continuous changes in user interests, and the coexistence of multiple behavioral optimization objectives. Learning large models in such an environment usually comes with higher training costs. Additionally, in scenarios requiring continuous training, it is essential to clarify the trade-off between short-term performance improvements and long-term stability. It remains unclear how model scale, update frequency, and memory budget jointly affect the long-term adaptive capability and forgetting degree of models.

\textbf{Token-to-item decoder for better diversity and multi-task performances.}
There still exists a semantic gap between current token-level generation and item-level output. Such token-to-item decoders can better align token-level generation with item-level semantics, and explicitly integrate diversity and multi-task optimization objectives in the decoding process. This helps alleviate the misalignment between generated content and final recommendation results, and meanwhile improves behavioral coverage and ranking quality.

\textbf{Dynamic user pool.}
In real-world systems, users constantly join, leave, and shift their interests over time. Modeling such user-side dynamics together with evolving item pools would provide a more realistic setting for studying adaptation, robustness, and long-term personalization.

\textbf{Low-activity and cold-start users.}  In this benchmark, we set user sequence lengths with $n\geq 100$. In the continual learning setting, this means that the day-averaged lower bound is around 3 interactions per user per day. The resulting data accounts for more than 93\% of the original log dataset before filtering. Yet, for extremely low-activity and cold-start users, though they are less representative in the industrial environment, they still account for users in millions in large-scale systems.
We considered it a different problem setting where continual learning on these users may require data collection across a much longer range, and these user groups require more careful modeling and evaluation methods.

\section{Broader Imapacts.}\label{app: borader_impacts}

Our work focuses on new-generation recommendation models and has both positive and negative implications for society. On the positive side, this publication will boost the research field forward to establish more robust, task-balanced, fairness-aware, and diversity-aware generative sequential recommendation models, and benefit the study of large models in general. The benchmark also potentially improves user engagement and satisfaction on web services. On the negative side, malicious use of this dataset may also exist, including attacks against recommender systems. Further complementary research in this direction may also be necessary, especially for large-scale datasets like ours.

\section{Detailed Diversity Metrics}\label{app: detail_diversity}

\begin{itemize}[leftmargin=*]
    \item \textbf{Gini Coefficient}: Measures the statistical inequality of token distribution across all generated SIDs, computed as the Gini coefficient of token frequencies.
    
    \item \textbf{TopK Layer Diversity}: Measures the diversity of tokens at each layer when evaluating top-K recommendations. For each layer $i$ in the SID, we compute the proportion of unique tokens among the top-K predictions.
    
    \item \textbf{Average Path Diversity}: Measures the average number of distinct paths (complete SIDs) from the root to layer $i$, reflecting the diversity of recommendation paths up to that layer.
    
    \item \textbf{Average Layer Unique Tokens}: When generating $K$ candidate SIDs, measures the average number of unique tokens at each layer across all $K$ candidates.
    
    \item \textbf{Average Layer Entropy}: Measures the entropy of token distribution at each layer when generating $K$ candidate SIDs, quantifying the uncertainty and diversity of predictions at each hierarchical level.
\end{itemize}

\section{Detailed Experiments Settings}\label{app:exp_setting}

\subsection{Models}\label{app:exp_models}

The models used in our experiments are as follows: 
\begin{itemize}[leftmargin=*]
    \setlength{\itemsep}{1pt}
    \setlength{\parskip}{1pt}
    \setlength{\topsep}{1pt}
\item \textbf{SASRec}\cite{kang2018selfattentivesequentialrecommendation} utilizes a self-attentive model to capture item correlations.
\item \textbf{GRURec}\cite{hidasi2016sessionbasedrecommendationsrecurrentneural} uses a Gated Recurring Unit (GRU) to enhance long-term memory in recommendation tasks.
\item \textbf{TIGER}\cite{rajput2023recommendersystemsgenerativeretrieval} proposes a novel learned Semantic ID representation for each item and achieved excellent performance in sequential recommendation tasks. In this paper, we use two sizes of the TIGER model.
\item \textbf{LLaDA-Rec}\cite{shi2025lladarecdiscretediffusionparallel} is a diffusion-based generative recommendation model that captures user preferences through iterative denoising for recommendation generation.
\item \textbf{MiniOneRec}\cite{kong2025minionerecopensourceframeworkscaling} is a lightweight open-sourced generative recommendation model that formulates recommendation as an autoregressive next-item generation task based on user interaction sequences.
\end{itemize}

\subsection{Experiments Settings}\label{app:exp_settings}
\textbf{SASRec} and \textbf{GRURec.}
The optimizer is AdamW, the learning rate is denoted by lr, weight decay by l2\_decay, the learning rate scheduler is get\_cosine\_schedule\_with\_warmup, the warm-up steps are warmup\_steps, and the loss function is BCEWithLogitsLoss. A set of basic hyperparameters is shared: epoch=10, batch\_size=512, hidden\_factor=256, lr=0.001, l2\_decay=0.05, warmup\_steps=1000, dropout\_rate=0.5, max\_k=200, patience=5, num\_negatives=1000, num\_workers=4, item\_num=2452987.

\textbf{TIGER.}
Both the Small and Tiny versions of the model adopt the official parameter sizes. We use GPUs and enable bf16 mixed precision. The core hyperparameters include: learning rate lr=1e-4, epochs per round epochs\_per\_round=1, training batch size train\_batch\_size=256, and evaluation batch size eval\_batch\_size=32.

\textbf{LLaDA-Rec.}
The model adopt the official parameter sizes. We use GPUs and enable bf16 mixed precision. The core hyperparameters include: learning rate lr=1e-3, epochs per round epochs\_per\_round=1, training batch size train\_batch\_size=256, and evaluation batch size eval\_batch\_size=32.

\textbf{MiniOneRec.}
The SFT experiment adopts Qwen2.5-0.5B as the base model, performs daily continued fine-tuning in streaming scenarios, and conducts GPU parallel training via accelerate launch. The training hyperparameters include num epochs = 1 and learning rate = 1e-4. The inference batch size is set to batch size = 4, and the beam search width is set to num beams = 50. For the overall experiment, only the click behavior click is retained as the behavior type for training and evaluation.

The RL experiment adopts SFT-Qwen2.5-0.5B as the base model and conducts continued training under the streaming reinforcement learning setting. The hyperparameters in the training phase are as follows: num train\_epochs=1, learning\_rate=1e-6, train\_batch\_size=256, rl eval batch size=32, gradient\_accumulation\_steps=1, temperature=1.0, num generations=8, beta=0.2. The adopted reward type is reward\_type=exact\_prefix\_behavior, which is introduced in Appendix \ref{app:exp_rl_reward}. Alternative options include exact, prefix, semantic, and exact\_prefix\_semantic. The weight and ratio parameters related to sample construction are w\_exact=5, w\_prefix=1, w\_sem=0.2, subtree sample ratio=0.02, w\_pos=10, w\_inter=10, w\_neg=1. Among them, the positive behavior is long\_view, interactive behaviors include like, follow, forward, comment, search and is\_profile\_enter, and the negative behavior is hate. The training also supports several toggle options, including add\_gt, beam\_search, test\_during\_training, dynamic\_sampling, sync\_ref\_model, dapo and gspo. The evaluation phase is executed automatically by default, adopting distributed inference with eval\_num\_gpus=8, along with eval\_batch\_size=16, eval\_num\_beams=50, eval\_max\_new\_tokens=32, eval\_k=50, and the random seed is set to seed=42.

\subsection{RL Reward Design}\label{app:exp_rl_reward}

While the SFT stage uses only click sequences, we design RL reward functions based on diverse behaviors: long-view, like, forward, comment, search, profile-enter, hate. More details about experiments setting and these behaviors can be found in appendix \ref{app:exp_setting}. As described in Section \ref{sec: benchmark_training}, we design the interaction reward as 
$$R_B = \sum_{b \in B} \alpha_b R_b.$$ 
Specifically, we regard long-view interactions as positive behaviors, while likes, forwards, comments, searches, and profile entries are defined as general interactive behaviors, and hate interactions are treated as negative behaviors. 

We also introduce additional reward designs to engage fine-grained optimization, including the SID prefix matching reward $$R_\text{pre}=\frac{1}{L} \sum_{i=1}^{L} \mathbb{I}[\text{prefix}_{i}^\text{gen}=\text{prefix}_{i}^\text{target}],$$ where $L$ is the total length of the SID. Moreover, we adopt a semantic reward $$R_\text{sem}=\text{cos}(\mathbf{e}_\text{gen}, \mathbf{e}_\text{target}),$$ determined by the cosine similarity between the generated SID and the target SID. 

The final reward is $$R =\lambda_{B} R_B + \lambda_\text{pre} R_\text{pre} + \lambda_\text{sem} R_\text{sem}.$$

\section{Data Specification}\label{app: data_description}

\subsection{Data Structure}

Each interaction sample contains the following fields:
\begin{itemize}[leftmargin=*]
    \item \textbf{user\_id}: an integer identifier of the user.
    \item \textbf{item\_id}: an integer identifier of the item, which is aligned with the item metadata file.
    \item \textbf{timestamp}: the interaction time in Unix epoch milliseconds, \eg, 1755009906758.
    \item \textbf{play-time} and \textbf{duration}: continuous feedback signals measured in milliseconds.
    \item \textbf{long-view}, \textbf{like}, \textbf{follow}, \textbf{forward}, \textbf{comment}, \textbf{search}, and \textbf{profile-enter}: binary positive labels indicating whether the corresponding behavior occurred.
    \item \textbf{hate}: a binary negative label indicating whether the user explicitly expressed negative feedback.
\end{itemize}

The item metadata file contains the following fields:
\begin{itemize}[leftmargin=*]
    \item \textbf{sid\_list}: the SID sequence representation of the item.
    \item \textbf{emb\_vec}: a 64-dimensional float vector, with each dimension ranging from $-16.0$ to $16.0$.
    \item \textbf{item\_id}: the item identifier, which corresponds to the \textbf{item\_id} in the interaction file.
    \item \textbf{tag\_list}: the item tags. For privacy protection, the original tags are anonymized and encoded as integers.
\end{itemize}

\subsection{Data Analysis}

\begin{table}[t]
    \centering
    \caption{Frequency quantiles of tags in KuaiRecV2.}
    \label{tab:tag_quantiles}
    \begin{tabular}{ccccccc}
        \toprule
        25\% & 50\% & 75\% & 90\% & 95\% & 99\% \\
        \midrule
        1.00 & 1.00 & 2.00 & 5.00 & 12.00 & 96.00 \\
        \bottomrule
    \end{tabular}
\end{table}

\begin{table}[t]
    \centering
    \caption{Number and proportion of tags exceeding different frequency thresholds in KuaiRecV2.}
    \label{tab:tag_thresholds}
    \begin{tabular}{ccc}
        \toprule
        Frequency Threshold & Number of Tags & Proportion \\
        \midrule
        $\geq 1$ & 11,068,844 & 100.00\% \\
        $\geq 5$ & 1,234,847 & 11.16\% \\
        $\geq 10$ & 677,351 & 6.12\% \\
        $\geq 50$ & 186,948 & 1.69\% \\
        $\geq 100$ & 107,387 & 0.97\% \\
        $\geq 500$ & 27,884 & 0.25\% \\
        $\geq 1{,}000$ & 15,123 & 0.14\% \\
        $\geq 5{,}000$ & 3,351 & 0.03\% \\
        $\geq 10{,}000$ & 1,619 & 0.01\% \\
        $\geq 50{,}000$ & 293 & 0.00\% \\
        $\geq 100{,}000$ & 120 & 0.00\% \\
        $\geq 500{,}000$ & 8 & 0.00\% \\
        $\geq 1{,}000{,}000$ & 2 & 0.00\% \\
        \bottomrule
    \end{tabular}
\end{table}

As show in Table \ref{tab:tag_quantiles} and Table \ref{tab:tag_thresholds}, the tag distribution of KuaiRecV2 is extremely long-tailed. Although the dataset contains 11.07M unique tags and 161.52M total tag occurrences, most tags are very sparse. The median frequency is only 1, and both the 25th and 50th percentiles are 1, which means that more than half of the tags appear only once. Even at the 90th percentile, a tag appears only 5 times, showing that the vast majority of tags are low-frequency. In contrast, only 0.97\% of tags appear at least 100 times, and only 8 tags appear at least 500,000 times. This highly skewed distribution indicates that KuaiRecV2 provides rich fine-grained semantic signals, while also posing substantial challenges for modeling long-tail and sparse tag information.

\begin{figure}[htbp]
    \centering
    \includegraphics[width=\linewidth]{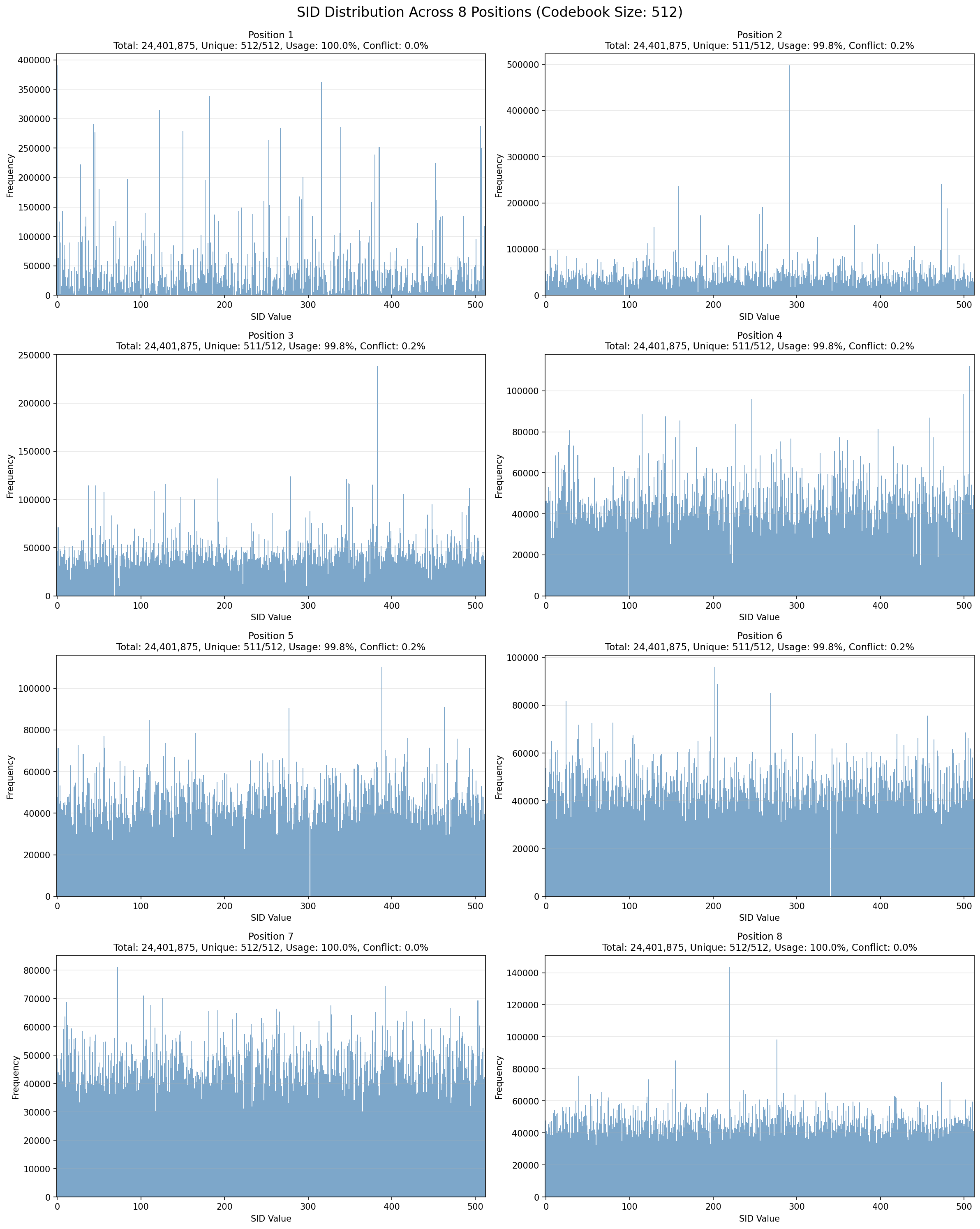}
    \caption{The distribution of SID, the SID distribution of higher-level codebooks is more uneven.}
    \label{fig: sid_distribution}
\end{figure}

SID is a hierarchical discrete representation with an 8-level structure and a codebook size of 512 at each level. The item metadata therefore, provides both discrete semantic codes and continuous embeddings, enabling richer modeling of item semantics and user-item interactions. The distribution of SID is show in figure \ref{fig: sid_distribution}. Our dataset contains a total of 24,401,875 items and 24,205,418 unique SID sequences, with an SID collision rate of 0.81\%.

\newpage

\section{Detailed results}  \label{app: detail_results}

\subsection{Comparison of results from different training data}\label{app: compare_trainingdata}

We use MiniOneRec to train on different behavior sequences and predict the long view target on day 21. The results are shown in the table \ref{tab:compare_trainingdata}, and training on a specific sequence can improve the prediction performance of that sequence.

\begin{table}[htbp]
  \centering
  \setlength{\tabcolsep}{4pt}
  \caption{Comparison of training and predicting long view results using click and long view sequences.}
  \label{tab:compare_trainingdata}
  \begin{tabular}{llcccccccccc}
    \toprule
    \multirow{2}{*}{Data} & \multirow{2}{*}{Behavior} & \multicolumn{5}{c}{NDCG ($\times 10^{-3}$)} &\multicolumn{5}{c}{HR ($\times 10^{-3}$)}\\
    \cmidrule(lr){3-7} \cmidrule(lr){8-12}
    & & @1 & @5 & @10 & @20 & @50 & @1 & @5 & @10 & @20 & @50\\
    \midrule
    click &long view & 3.18 & 8.02 & 10.78 & 13.97 & 17.43& 3.18 & 12.89 & 21.49 & 34.20 & 51.50 \\
    long view &long view & 3.50 & 8.90 & 11.89 & 15.33 & 19.76& 3.50 & 14.37 & 23.72 & 37.40 & 59.60 \\
    \bottomrule
  \end{tabular}
\end{table}

\subsection{Detailed Analysis of Token-Level Diversity}\label{app: token_diversity_case_study}
In this section, we take 50 responses generated by one model for a single sequence as an example. We use the same color to mark identical tokens at this level. We find that when the prefix tokens are the same or similar, the model is more likely to generate the same next token, and may even produce nearly identical SID combinations. Therefore, the token level diversity metric can reflect in more detail that the diversity of the recommended items generated by the model is poor.

\begin{Verbatim}[
    frame=single, % 加边框形成整块
    fontsize=\small,
    samepage=false, % 核心：关闭同页限制，允许跨页
    xleftmargin=1em,
    xrightmargin=1em,
    commandchars=\\\{\}, % 允许内部使用颜色命令
    numbers=left, 
    firstnumber=1, 
    stepnumber=1,
    numbersep=5pt,
    numberblanklines=false,
    formatcom=\linespread{0.8}\selectfont
    ]
\textcolor{red}{<a\_181>}\textcolor{green}{<b\_183>}\textcolor{cyan}{<c\_348>}<d\_31><e\_26><f\_140><g\_297><h\_67>
\textcolor{red}{<a\_181>}\textcolor{green}{<b\_183>}\textcolor{cyan}{<c\_348>}<d\_243><e\_448><f\_319><g\_35><h\_383>
\textcolor{red}{<a\_181>}\textcolor{green}{<b\_183>}\textcolor{blue}{<c\_387>}\textcolor{yellow}{<d\_168><e\_251><f\_140><g\_105>}<h\_336>
<a\_10><b\_85><c\_493><d\_67><e\_1><f\_47><g\_371><h\_428>
<a\_200><b\_115><c\_76><d\_349><e\_269><f\_334><g\_478><h\_119>
<a\_10><b\_158><c\_341><d\_15><e\_20><f\_269><g\_498><h\_498>
\textcolor{red}{<a\_181>}<b\_198><c\_383><d\_250><e\_258><f\_421><g\_484><h\_429>
\textcolor{red}{<a\_181>}\textcolor{green}{<b\_183>}<c\_394><d\_431><e\_251><f\_376><g\_431><h\_375>
\textcolor{red}{<a\_181>}<b\_224><c\_382><d\_451><e\_368><f\_342><g\_304><h\_116>
<a\_10><b\_400><c\_214><d\_34><e\_56><f\_362><g\_331><h\_42>
\textcolor{red}{<a\_181>}\textcolor{green}{<b\_183>}\textcolor{cyan}{<c\_348>}<d\_420><e\_7><f\_6><g\_410><h\_69>
\textcolor{red}{<a\_181>}\textcolor{green}{<b\_183>}\textcolor{cyan}{<c\_348>}<d\_443><e\_13><f\_14><g\_477><h\_75>
<a\_129><b\_394><c\_107><d\_368><e\_254><f\_49><g\_260><h\_54>
<a\_500><b\_228><c\_191><d\_132><e\_96><f\_18><g\_191><h\_455>
\textcolor{red}{<a\_181>}<b\_198><c\_350><d\_439><e\_25><f\_20><g\_503><h\_376>
<a\_315><b\_179><c\_146><d\_310><e\_96><f\_19><g\_190><h\_181>
\textcolor{red}{<a\_181>}<b\_480><c\_334><d\_483><e\_132><f\_187><g\_240><h\_471>
\textcolor{red}{<a\_181>}<b\_198><c\_414><d\_142><e\_343><f\_491><g\_5><h\_39>
<a\_453><b\_291><c\_182><d\_483><e\_396><f\_448><g\_369><h\_120>
<a\_500><b\_343><c\_340><d\_247><e\_481><f\_166><g\_351><h\_424>
<a\_163><b\_168><c\_209><d\_106><e\_60><f\_231><g\_263><h\_50>
<a\_405><b\_291><c\_382><d\_1><e\_212><f\_360><g\_96><h\_276>
\textcolor{red}{<a\_181>}\textcolor{green}{<b\_183>}\textcolor{cyan}{<c\_348>}<d\_403><e\_311><f\_267><g\_270><h\_494>
<a\_453><b\_291><c\_471><d\_318><e\_98><f\_111><g\_288><h\_315>
<a\_203><b\_291><c\_302><d\_449><e\_437><f\_285><g\_469><h\_286>
\textcolor{red}{<a\_181>}<b\_226><c\_257><d\_407><e\_287><f\_112><g\_9><h\_193>
\textcolor{red}{<a\_181>}<b\_88><c\_192><d\_361><e\_15><f\_200><g\_103><h\_62>
<a\_10><b\_158><c\_279><d\_207><e\_478><f\_300><g\_112><h\_391>
\textcolor{red}{<a\_181>}\textcolor{green}{<b\_183>}<c\_218><d\_106><e\_419><f\_238><g\_452><h\_384>
<a\_500><b\_228><c\_505><d\_312><e\_196><f\_18><g\_10><h\_414>
\textcolor{red}{<a\_181>}\textcolor{green}{<b\_183>}\textcolor{cyan}{<c\_348>}<d\_131><e\_225><f\_140><g\_480><h\_169>
\textcolor{red}{<a\_181>}<b\_224><c\_199><d\_115><e\_228><f\_57><g\_498><h\_331>
<a\_453><b\_291><c\_81><d\_290><e\_230><f\_130><g\_460><h\_276>
<a\_129><b\_291><c\_458><d\_251><e\_363><f\_54><g\_41><h\_450>
<a\_453><b\_291><c\_334><d\_429><e\_355><f\_421><g\_327><h\_315>
\textcolor{red}{<a\_181>}<b\_205><c\_429><d\_214><e\_395><f\_488><g\_83><h\_490>
<a\_10><b\_291><c\_410><d\_382><e\_493><f\_277><g\_274><h\_473>
\textcolor{red}{<a\_181>}<b\_224><c\_383><d\_503><e\_507><f\_437><g\_440><h\_450>
\textcolor{red}{<a\_181>}\textcolor{green}{<b\_183>}\textcolor{cyan}{<c\_348>}<d\_302><e\_379><f\_485><g\_470><h\_383>
\textcolor{red}{<a\_181>}\textcolor{green}{<b\_183>}\textcolor{blue}{<c\_387>}\textcolor{yellow}{<d\_168><e\_251><f\_140><g\_105>}<h\_138>
\textcolor{red}{<a\_181>}\textcolor{green}{<b\_183>}<c\_148><d\_178><e\_53><f\_351><g\_5><h\_80>
\textcolor{red}{<a\_181>}\textcolor{green}{<b\_183>}\textcolor{blue}{<c\_387>}\textcolor{yellow}{<d\_168><e\_251><f\_140><g\_105>}<h\_439>
\textcolor{red}{<a\_181>}\textcolor{green}{<b\_183>}\textcolor{blue}{<c\_387>}\textcolor{yellow}{<d\_168><e\_251><f\_140><g\_105>}<h\_124>
\textcolor{red}{<a\_181>}\textcolor{green}{<b\_183>}\textcolor{cyan}{<c\_348>}<d\_131><e\_446><f\_212><g\_480><h\_227>
\textcolor{red}{<a\_181>}\textcolor{green}{<b\_183>}\textcolor{blue}{<c\_387>}\textcolor{yellow}{<d\_168><e\_251><f\_140><g\_105>}<h\_253>
\textcolor{red}{<a\_181>}\textcolor{green}{<b\_183>}\textcolor{blue}{<c\_387>}\textcolor{yellow}{<d\_168><e\_251><f\_140><g\_105>}<h\_254>
\textcolor{red}{<a\_181>}\textcolor{green}{<b\_183>}\textcolor{blue}{<c\_387>}\textcolor{yellow}{<d\_168><e\_251><f\_140><g\_105>}<h\_48>
\textcolor{red}{<a\_181>}\textcolor{green}{<b\_183>}\textcolor{blue}{<c\_387>}\textcolor{yellow}{<d\_168><e\_251><f\_140><g\_105>}<h\_128>
\textcolor{red}{<a\_181>}\textcolor{green}{<b\_183>}<c\_394><d\_431><e\_251><f\_376><g\_431><h\_138>
<a\_200><b\_115><c\_76><d\_349><e\_269><f\_334><g\_478><h\_48>
\end{Verbatim}

\subsection{Detailed Behavior Metrics Across Streaming Training Days} \label{app: behavior_results}

This section presents comprehensive behavior-specific metrics of MiniOneRec as the example for continual learning of large generative recommenders.
We show results in Tables for day 1, day 21, and day 30.
For each training day, we report NDCG (Normalized Discounted Cumulative Gain) and HR (Hit Rate) at different cutoff values (@1, @5, @10, @20, @50) for various user behaviors. The behaviors include overall performance as well as specific interactions such as long view, like, follow, forward, comment, search, profile enter, and hate. These detailed results demonstrate how recommendation quality evolves over time for different types of user engagement. All NDCG and HR values are scaled by $10^{-3}$ for readability. The variation in metrics across behaviors reveals the model's ability to capture diverse user intents, while the progression across rounds illustrates the adaptation dynamics in the streaming recommendation scenario.

\begin{table}[htbp]
  \centering
  \caption{Behavior metrics at different cutoffs for streaming training day 1.}
  \label{tab:behavior_metrics_round_0}
  \begin{tabular}{lcccccccccc}
    \toprule
    \multirow{2}{*}{Behavior} & \multicolumn{5}{c}{NDCG ($\times 10^{-3}$)} &\multicolumn{5}{c}{HR ($\times 10^{-3}$)}\\
    \cmidrule(lr){2-6} \cmidrule(lr){7-11}
    & @1 & @5 & @10 & @20 & @50 & @1 & @5 & @10 & @20 & @50\\
    \midrule
    overall & 0.34 & 2.14 & 3.60 & 5.63 & 8.44 & 0.34 & 4.13 & 8.68 & 16.73 & 30.92 \\
    long view & 0.35 & 2.27 & 3.85 & 6.04 & 9.16 & 0.35 & 4.41 & 9.34 & 18.03 & 33.82 \\
    like & 0.66 & 2.28 & 3.45 & 5.29 & 8.66 & 0.66 & 3.97 & 7.61 & 14.89 & 32.09 \\
    follow & 0.00 & 2.55 & 2.55 & 5.12 & 7.60 & 0.00 & 6.60 & 6.60 & 16.50 & 29.70 \\
    forward & 0.00 & 2.46 & 2.46 & 4.28 & 5.29 & 0.00 & 4.93 & 4.93 & 12.32 & 17.24 \\
    comment & 0.94 & 2.80 & 4.92 & 7.72 & 11.21 & 0.94 & 4.69 & 11.25 & 22.49 & 40.30 \\
    hate & 0.00 & 0.00 & 0.00 & 0.00 & 0.00 & 0.00 & 0.00 & 0.00 & 0.00 & 0.00 \\
    search & 0.00 & 0.00 & 0.97 & 1.74 & 3.64 & 0.00 & 0.00 & 3.22 & 6.43 & 16.08 \\
    profile enter & 0.32 & 3.04 & 4.12 & 6.22 & 8.42 & 0.32 & 6.12 & 9.34 & 17.71 & 28.65 \\
    \bottomrule
  \end{tabular}
\end{table}

\begin{table}[htbp]
  \centering
  \caption{Behavior metrics at different cutoffs for streaming training day 21.}
  \label{tab:behavior_metrics_round_21}
  \begin{tabular}{lcccccccccc}
    \toprule
    \multirow{2}{*}{Behavior} & \multicolumn{5}{c}{NDCG ($\times 10^{-3}$)} &\multicolumn{5}{c}{HR ($\times 10^{-3}$)}\\
    \cmidrule(lr){2-6} \cmidrule(lr){7-11}
    & @1 & @5 & @10 & @20 & @50 & @1 & @5 & @10 & @20 & @50\\
    \midrule
    overall & 2.87 & 7.30 & 9.85 & 12.78 & 16.03 & 2.87 & 11.78 & 19.70 & 31.40 & 47.60 \\
    long view & 3.18 & 8.02 & 10.78 & 13.97 & 17.43 & 3.18 & 12.90 & 21.50 & 34.21 & 51.50 \\
    like & 2.48 & 7.05 & 9.48 & 12.47 & 16.45 & 2.48 & 12.04 & 19.48 & 31.52 & 50.99 \\
    follow & 0.00 & 7.30 & 9.66 & 10.53 & 10.53 & 0.00 & 15.33 & 22.99 & 26.82 & 26.82 \\
    forward & 2.64 & 5.44 & 6.32 & 8.30 & 9.82 & 2.64 & 7.92 & 10.55 & 18.47 & 26.39 \\
    comment & 3.74 & 5.20 & 7.29 & 9.38 & 12.72 & 3.74 & 6.54 & 13.08 & 21.50 & 38.32 \\
    hate & 0.00 & 0.00 & 0.00 & 0.00 & 0.00 & 0.00 & 0.00 & 0.00 & 0.00 & 0.00 \\
    search & 0.00 & 3.26 & 10.71 & 14.63 & 16.18 & 0.00 & 7.97 & 31.87 & 47.81 & 55.78 \\
    profile enter & 2.71 & 5.20 & 7.70 & 10.24 & 13.09 & 2.71 & 8.13 & 15.92 & 26.08 & 40.30 \\
    \bottomrule
  \end{tabular}
\end{table}

\begin{table}[htbp]
  \centering
  \caption{Behavior metrics at different cutoffs for streaming training day 30.}
  \label{tab:behavior_metrics_round_30}
  \begin{tabular}{lcccccccccc}
    \toprule
    \multirow{2}{*}{Behavior} & \multicolumn{5}{c}{NDCG ($\times 10^{-3}$)} &\multicolumn{5}{c}{HR ($\times 10^{-3}$)}\\
    \cmidrule(lr){2-6} \cmidrule(lr){7-11}
    & @1 & @5 & @10 & @20 & @50 & @1 & @5 & @10 & @20 & @50\\
    \midrule
    overall & 2.24 & 5.91 & 8.09 & 10.67 & 13.69 & 2.24 & 9.70 & 16.47 & 26.73 & 41.83 \\
    long view & 2.45 & 6.38 & 8.80 & 11.55 & 14.80 & 2.45 & 10.40 & 17.93 & 28.87 & 45.11 \\
    like & 0.74 & 3.69 & 5.45 & 7.88 & 9.77 & 0.74 & 6.64 & 12.17 & 21.76 & 31.35 \\
    follow & 0.00 & 1.79 & 1.79 & 2.88 & 3.63 & 0.00 & 4.15 & 4.15 & 8.30 & 12.45 \\
    forward & 2.75 & 4.13 & 7.55 & 9.69 & 10.82 & 2.75 & 5.51 & 16.53 & 24.79 & 30.30 \\
    comment & 0.98 & 4.28 & 6.49 & 7.48 & 10.55 & 0.98 & 7.82 & 14.66 & 18.57 & 34.21 \\
    hate & 0.00 & 19.12 & 19.12 & 19.12 & 25.29 & 0.00 & 30.30 & 30.30 & 30.30 & 60.61 \\
    search & 12.24 & 14.29 & 15.51 & 20.39 & 21.28 & 12.24 & 16.33 & 20.41 & 40.82 & 44.90 \\
    profile enter & 2.28 & 5.76 & 7.46 & 9.36 & 11.68 & 2.28 & 9.44 & 14.64 & 22.13 & 33.52 \\
    \bottomrule
  \end{tabular}
\end{table}

\newpage

\subsection{Layer-wise Diversity Metrics Across Streaming Rounds}\label{app: diversity_results}

This section provides layer-wise diversity analysis of MiniOneRec for the click behavior on streaming training day 1, 21, 30. Details of these metrics are provided in Appendix \ref{app: detail_diversity}. 
\begin{table}[htbp]
  \centering
  \caption{Diversity metrics across layers for streaming training day 1 (click behavior).}
  \label{tab:diversity_metrics_round_0}
  \begin{tabular}{llcccccccc}
    \toprule
    \multicolumn{2}{c}{Metric} & \multicolumn{8}{c}{Layer}\\
    \cmidrule(lr){3-10}
    & & 1 & 2 & 3 & 4 & 5 & 6 & 7 & 8\\
    \midrule
    \multicolumn{2}{l}{Avg Path Diversity} & 11.02 & 32.58 & 46.23 & 49.09 & 49.58 & 49.89 & 49.94 & 50.00 \\
    \multicolumn{2}{l}{Avg Layer Unique Tokens} & 11.02 & 26.76 & 41.83 & 44.96 & 46.08 & 46.96 & 46.77 & 47.23 \\
    \multicolumn{2}{l}{Avg Layer Entropy} & 2.47 & 4.21 & 5.26 & 5.43 & 5.48 & 5.52 & 5.51 & 5.53 \\
    \midrule
    \multirow{5}{*}{\begin{tabular}[c]{@{}l@{}}TopK Layer\\Diversity\end{tabular}} & (Top1) & 1.00 & 1.00 & 1.00 & 1.00 & 1.00 & 1.00 & 1.00 & 1.00 \\
     & (Top5) & 3.32 & 4.67 & 4.93 & 4.96 & 4.98 & 4.98 & 4.97 & 4.98 \\
     & (Top10) & 5.23 & 8.84 & 9.71 & 9.82 & 9.89 & 9.90 & 9.87 & 9.90 \\
     & (Top20) & 7.59 & 15.74 & 18.91 & 19.29 & 19.47 & 19.55 & 19.47 & 19.59 \\
     & (Top50) & 11.02 & 26.76 & 41.83 & 44.96 & 46.08 & 46.96 & 46.77 & 47.23 \\
    \bottomrule
  \end{tabular}
\end{table}

\begin{table}[htbp]
  \centering
  \caption{Diversity metrics across layers for streaming training day 21 (click behavior).}
  \label{tab:diversity_metrics_round_20}
  \begin{tabular}{llcccccccc}
    \toprule
    \multicolumn{2}{c}{Metric} & \multicolumn{8}{c}{Layer}\\
    \cmidrule(lr){3-10}
    & & 1 & 2 & 3 & 4 & 5 & 6 & 7 & 8\\
    \midrule
    \multicolumn{2}{l}{Avg Path Diversity} & 12.90 & 34.85 & 46.50 & 49.26 & 49.73 & 49.87 & 49.95 & 50.00 \\
    \multicolumn{2}{l}{Avg Layer Unique Tokens} & 12.90 & 28.62 & 41.81 & 45.40 & 45.99 & 46.50 & 47.29 & 47.27 \\
    \multicolumn{2}{l}{Avg Layer Entropy} & 2.75 & 4.35 & 5.26 & 5.45 & 5.47 & 5.50 & 5.53 & 5.53 \\
    \midrule
    \multirow{5}{*}{\begin{tabular}[c]{@{}l@{}}TopK Layer\\Diversity\end{tabular}} & (Top1) & 1.00 & 1.00 & 1.00 & 1.00 & 1.00 & 1.00 & 1.00 & 1.00 \\
     & (Top5) & 3.59 & 4.76 & 4.89 & 4.96 & 4.96 & 4.97 & 4.98 & 4.99 \\
     & (Top10) & 5.89 & 9.13 & 9.65 & 9.86 & 9.85 & 9.88 & 9.90 & 9.92 \\
     & (Top20) & 8.93 & 16.63 & 18.85 & 19.41 & 19.41 & 19.50 & 19.59 & 19.63 \\
     & (Top50) & 12.90 & 28.62 & 41.81 & 45.40 & 45.99 & 46.50 & 47.29 & 47.27 \\
    \bottomrule
  \end{tabular}
\end{table}

\begin{table}[htbp]
  \centering
  \caption{Diversity metrics across layers for streaming training day 30 (click behavior).}
  \label{tab:diversity_metrics_round_30}
  \begin{tabular}{llcccccccc}
    \toprule
    \multicolumn{2}{c}{Metric} & \multicolumn{8}{c}{Layer}\\
    \cmidrule(lr){3-10}
    & & 1 & 2 & 3 & 4 & 5 & 6 & 7 & 8\\
    \midrule
    \multicolumn{2}{l}{Avg Path Diversity} & 12.48 & 34.55 & 45.94 & 48.77 & 49.68 & 49.86 & 49.92 & 50.00 \\
    \multicolumn{2}{l}{Avg Layer Unique Tokens} & 12.48 & 28.57 & 41.12 & 44.96 & 45.63 & 46.60 & 46.96 & 47.02 \\
    \multicolumn{2}{l}{Avg Layer Entropy} & 2.77 & 4.40 & 5.22 & 5.43 & 5.46 & 5.50 & 5.52 & 5.52 \\
    \midrule
    \multirow{5}{*}{\begin{tabular}[c]{@{}l@{}}TopK Layer\\Diversity\end{tabular}} & (Top1) & 1.00 & 1.00 & 1.00 & 1.00 & 1.00 & 1.00 & 1.00 & 1.00 \\
     & (Top5) & 3.55 & 4.77 & 4.91 & 4.96 & 4.92 & 4.98 & 4.98 & 4.98 \\
     & (Top10) & 5.73 & 9.14 & 9.63 & 9.82 & 9.76 & 9.89 & 9.90 & 9.91 \\
     & (Top20) & 8.62 & 16.61 & 18.71 & 19.30 & 19.25 & 19.51 & 19.56 & 19.59 \\
     & (Top50) & 12.48 & 28.57 & 41.12 & 44.96 & 45.63 & 46.60 & 46.96 & 47.02 \\
    \bottomrule
  \end{tabular}
\end{table}

\end{document}